\documentclass{article}

\usepackage{ijcai25}

\usepackage{times}
\usepackage{soul}
\usepackage{url}
\usepackage[hidelinks]{hyperref}
\usepackage[utf8]{inputenc}
\usepackage[small]{caption}
\usepackage{graphicx}
\usepackage{amsmath}
\usepackage{amsthm}
\usepackage{booktabs}
\usepackage[ruled,vlined,linesnumbered]{algorithm2e}
\usepackage[switch]{lineno}
\usepackage{fix-cm}

\usepackage{subfigure}
\usepackage[english]{babel}
\usepackage{blindtext}
\usepackage{subcaption}
\usepackage{listings}
\usepackage{xcolor}
\usepackage{balance}

\hypersetup{
    colorlinks=true,  
    linkcolor=black,
    filecolor=blue,      
    urlcolor=black,
    citecolor=black, 
}

\usepackage{array}
\newcolumntype{M}[1]{>{\centering\arraybackslash}p{#1}}

\usepackage{latexsym}

\title{EACO-RAG: Towards Distributed Tiered LLM Deployment using \\ Edge-Assisted and Collaborative RAG with Adaptive Knowledge Update}

\author{
    Anonymous Author(s)
}

\author{
    Jiaxing Li$^1$, Chi Xu$^1$, Lianchen Jia$^2$, Feng Wang$^3$, Cong Zhang$^4$, Jiangchuan Liu$^1$
    \affiliations
    $^1$Simon Fraser University, Burnaby, BC, Canada \space
    $^2$Tsinghua University, Beijing, China \\
    $^3$University of Mississippi, University, MS, USA \space
    $^4$Jiangxing Intelligence Inc., Shenzhen, China
    \emails
    \{jla641, chix, jcliu\}@sfu.ca, jlc21@mails.tsinghua.edu.cn, fwang@cs.olemiss.edu, vcongzc@gmail.com
}

\begin{document}
\maketitle

\begin{abstract}
Large language models (LLMs) have demonstrated impressive capabilities in language tasks, but they require high computing power and rely on static knowledge. To overcome these limitations, Retrieval-Augmented Generation (RAG) incorporates up-to-date external information into LLMs without extensive fine-tuning. Meanwhile, small language models (SLMs) deployed on edge devices offer efficiency and low latency but often struggle with complex reasoning tasks. Unfortunately, current RAG approaches are predominantly based on centralized databases and have not been adapted to address the distinct constraints associated with deploying SLMs in edge environments. To bridge this gap, we propose Edge-Assisted and Collaborative RAG (EACO-RAG), a lightweight framework that leverages distributed edge nodes for adaptive knowledge updates and retrieval. EACO-RAG also employs a hierarchical collaborative gating mechanism to dynamically select among local, edge-assisted, and cloud-based strategies, with a carefully designed algorithm based on Safe Online Bayesian Optimization to maximize the potential performance enhancements. Experimental results demonstrate that EACO-RAG matches the accuracy of cloud-based knowledge graph RAG systems while reducing total costs by up to 84.6\% under relaxed delay constraints and by 65.3\% under stricter delay requirements. This work represents our initial effort toward achieving a distributed and scalable tiered LLM deployments, with EACO-RAG serving as a promising first step in unlocking the full potential of hybrid edge–cloud intelligence.


\end{abstract}

\vspace{-0.35cm}
\section{Introduction}

In recent years, Large Language Models (LLMs) have made remarkable strides in natural language comprehension and generation, drawing significant interest from both academia and industry~\cite{chang2024survey,wei2022emergent,ijcai2024p921}. They are revolutionizing real-world applications by enabling more intelligent, adaptive, and scalable solutions. Their integration not only enhances user experiences with accurate, real-time, and context-aware responses but also improves system efficiency and unlocks new capabilities in autonomous processing~\cite{wu2022autoformalization}, recommendations~\cite{lyu2023llm,ren2024representation}, and decision-making~\cite{yang2023large,li2022pre}.

On the other hand, Retrieval-Augmented Generation (RAG) enhances LLMs by incorporating retrieval mechanisms from external knowledge bases. Although LLMs benefit from vast pre-trained knowledge, their static nature limits performance when information evolves rapidly. RAG addresses this limitation by providing up-to-date context during inference, which improves response accuracy. Modern LLMs scale to billions or even trillions of parameters, demanding extensive computing resources~\cite{ijcai2024p0705,guo2025deepseek}. Given the high cost of retraining, RAG offers an efficient alternative by dynamically integrating current knowledge. Consequently, RAG-based methods are increasingly adopted in healthcare, education, and legal services, with a projected compound annual growth rate of 44.7\% from 2024 to 2030~\cite{Grandview2024}.

In industrial deployments, there is a trend to shift toward solutions with lower latency and reduced hardware costs. This trend drives the adoption of small-scale LLMs, or small language models (SLMs), for edge and on-device applications~\cite{zhang2024edgeshard,qu2024mobile}. Nevertheless, SLMs often struggle with tasks that require extensive external knowledge, leading to higher hallucination rates~\cite{chen2024benchmarking,yang2018hotpotqa}. As such, paring RAG with SLMs can compensate for their limited capacity by retrieving relevant external information, and thus enhance factual accuracy and contextual relevance while safeguarding data privacy and reducing delay—features crucial for industrial applications with different Quality of Service (QoS) requirements.

However, current RAG solutions are not yet optimized for edge distributed deployments of SLMs. They often rely on centralized databases and face several key challenges. First, large-scale retrieval can introduce redundant information, leading to increased latency and inference costs~\cite{hofstatter2023fid,yu2024evaluation}. Second, irrelevant or misleading retrieval degrades output quality~\cite{chen2024benchmarking}. Third, complex or multi-hop queries may exceed the capacity of lightweight databases and models, necessitating deeper retrieval and reasoning methods~\cite{ijcai2024p0734}. Finally, most existing RAG systems cannot dynamically adapt to ever-changing user interests across regions and over time.

\begin{figure}[t]
\centering
\includegraphics[width=0.48\textwidth]{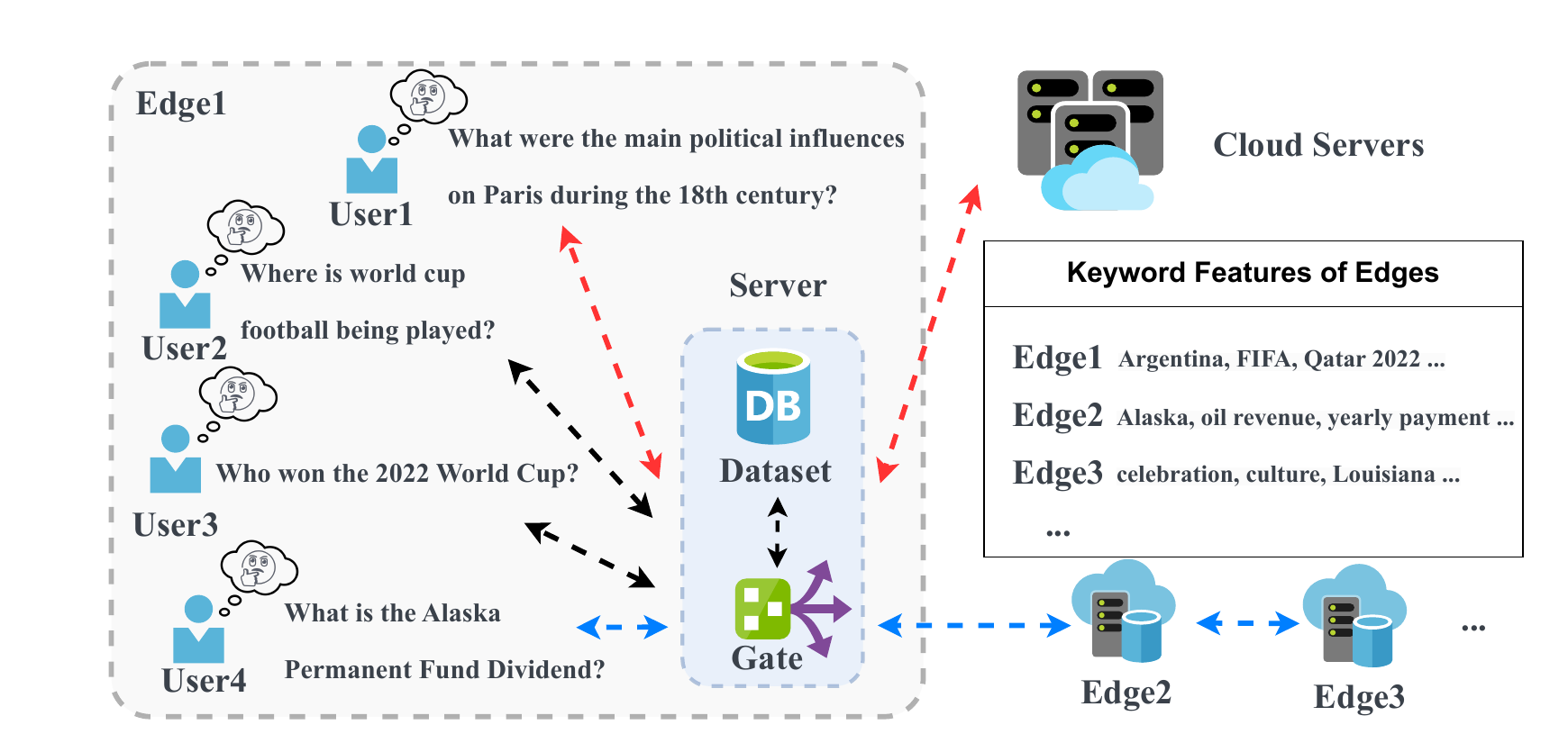}
\caption{EACO-RAG's adaptive retrieval. Each edge maintains a dynamic local dataset of popular topics. The collaborative gating mechanism selects retrieval sources from local, edge, or cloud datasets to adapt to evolving user interests and knowledge distributions. Black, blue, and red arrows denote local, edge-assisted, and cloud communications, respectively.}
\vspace{-0.35cm} 
\label{fig:EACO_RAG_Example}
\end{figure}

To address these limitations and facilitate highly distributed, scalable deployments, we propose Edge-Assisted and Collaborative RAG (EACO-RAG). Rather than relying on a single, monolithic knowledge base, EACO-RAG employs a distributed approach across multiple edge nodes, enabling broader and more contextually relevant retrieval beyond immediate local trends. Moreover, each node adaptively updates its local knowledge sets, ensuring timely accuracy in response to evolving information. A hierarchical collaborative gating mechanism is also designed to determine whether to perform retrieval and generation locally, via edge assistance, or in the cloud. This approach allows SLMs at the edge to work in tandem with powerful cloud resources, thereby reducing costs while maintaining accuracy and satisfying delay requirements. Figure~\ref{fig:EACO_RAG_Example} illustrates a toy example of how EACO-RAG adapts retrieval across different levels.

Through these optimizations, EACO-RAG substantially lowers inference costs and retrieval delays while maintaining high accuracy. Our experiments show that under relaxed delay constraints, EACO-RAG achieves accuracy comparable to cloud-based 72B LLM+GraphRAG while reducing cost by up to 84.6\%. Even when maintaining similar accuracy and delay to LLM+GraphRAG, it still reduces cost by up to 65.3\%. These results underscore the potential of EACO-RAG as a promising method toward distributed tiered LLM deployments, bridging edge and cloud to unlock more efficient, scalable, and adaptable AI systems.

The contributions of this work are summarized as follows:

\begin{itemize} 
    \item To the best of our knowledge, this paper is the first effort to systematically propose and investigate an edge-assisted distributed RAG architecture, which leverages adaptive knowledge updates and collaboration across edge nodes and cloud resources to offer a cost-efficient solution for large-scale distributed environments.
    \item We design dynamic update and edge-assisted distributed mechanisms that enable each edge node to adjust its local knowledge set. This ensures the real-time validity and effectiveness of edge data, also avoids the limitations of relying on a single dataset.
    \item We develop a context-aware collaborative gating mechanism that leverages Safe Online Bayesian Optimization for cost-effective decision-making and dynamic adaptation to evolving user demands and QoS constraints. 
    \item We conduct extensive experiments evaluating EACO-RAG, demonstrating its outperforms centralized RAG systems in terms of retrieval relevance, adaptability to evolving information, scalability to different scenarios and efficiency in balancing accuracy, delay, and cost.

\end{itemize}

\vspace{-0.3cm}
\section{Background and Motivation}

\begin{figure}[!t]
\small
  \centering
  \subfigure{
    \includegraphics[width=0.45\linewidth]{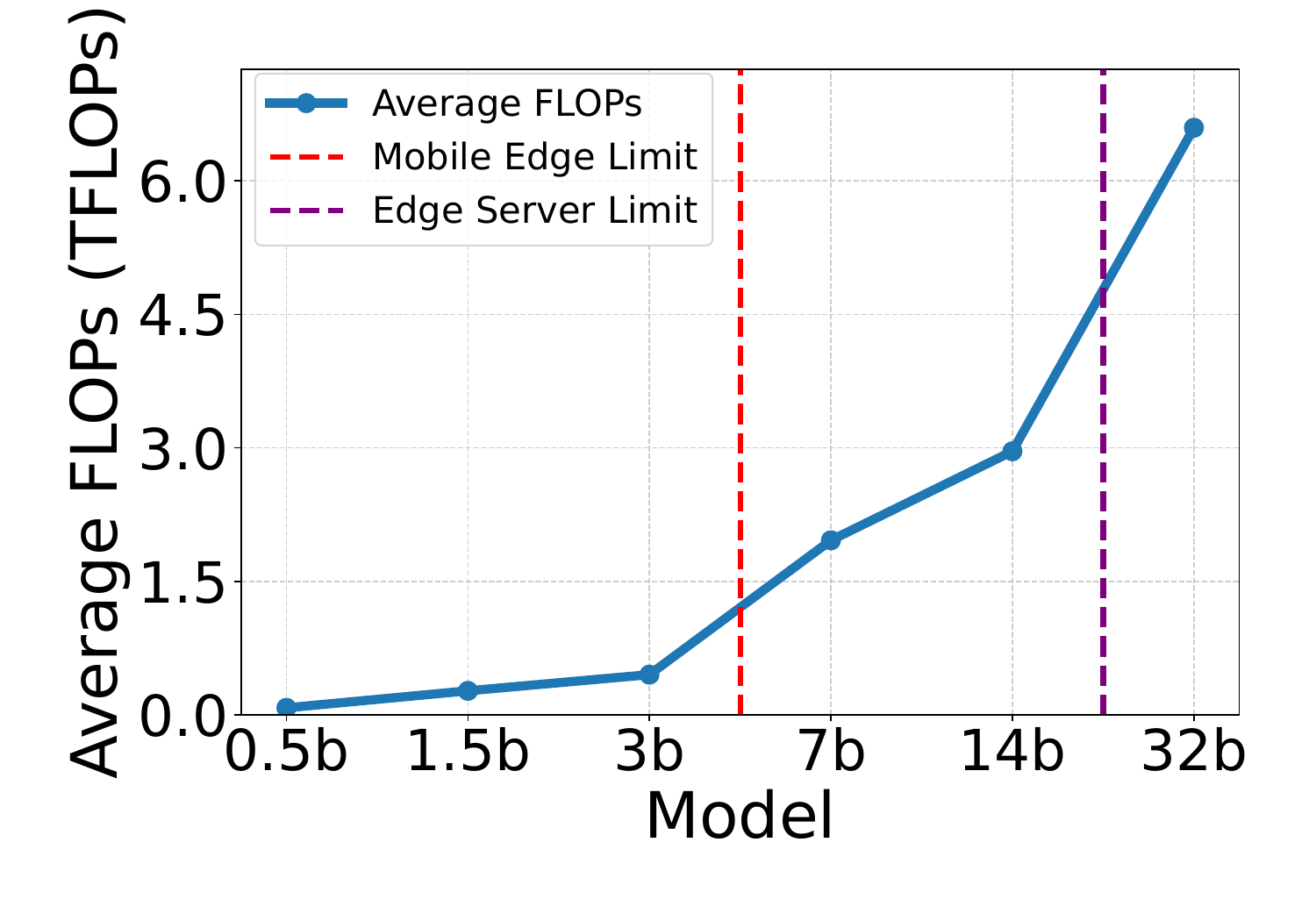}
    \vspace{-5pt}
    \label{Figures/MEASUREMENT/Figure1_models_flops_size.pdf}
  }
  \subfigure{
    \includegraphics[width=0.45\linewidth]{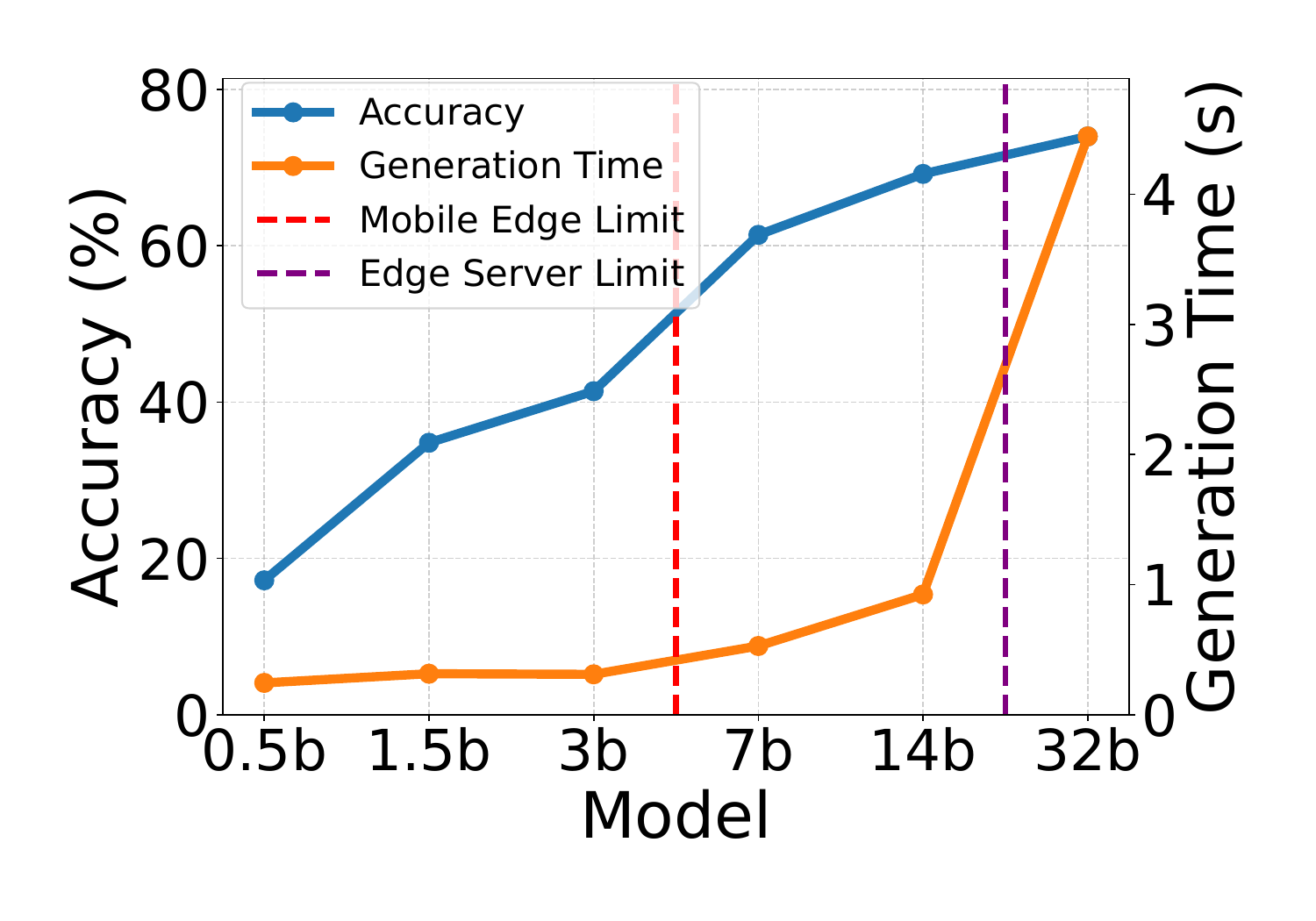}
    \vspace{-5pt}
    \label{Figures/MEASUREMENT/Figure1_models_accuracy_generatetime.pdf}
  }
  \vspace{-5pt}
  \captionsetup{width=1\linewidth} 
  \vspace{-5pt}
  \caption{Performance trade-offs in LLM-only applications using Qwen2.5~\protect\cite{qwen2}, evaluated on the TriviaQA dataset~\protect\cite{joshi2017triviaqa}. \textbf{Left:} Model size vs. inference cost, showing the relationship between LLM parameters and TFLOPs. \textbf{Right:} Model size vs. accuracy and delay, illustrating the impact of LLM  parameter on accuracy and generation latency.}

  \label{fig:Combined_Model_Performance_and_Bottleneck}
  \vspace{-0.18cm} 
\end{figure}

Recent advancements in LLMs have explored both scaling up models for enhanced reasoning and optimizing smaller models for efficiency~\cite{qwen2,guo2025deepseek}. Large-scale language models, such as OpenAI o1\footnote{\url{https://openai.com/o1/}}, o3\footnote{\url{https://openai.com/index/openai-o3-mini/}}, and DeepSeek R1\footnote{\url{https://api-docs.deepseek.com/news/news250120}}, push the boundaries of accuracy and complex reasoning by leveraging extensive computational resources. While these models achieve unprecedented performance, their high inference costs and delay make them less practical for real-time and resource-constrained applications.

\begin{table}[t]
\small
\centering
\begin{tabular}{p{1.3cm}p{1.8cm}p{1.8cm}p{1.9cm}}
\toprule
\textbf{{\fontsize{8.5pt}{10pt}\selectfont Approach}} & \textbf{{\fontsize{8.5pt}{10pt}\selectfont Input Token}} & \textbf{{\fontsize{8.5pt}{10pt}\selectfont Output Token}} & \textbf{{\fontsize{8.5pt}{10pt}\selectfont Inference Cost}}\\
\midrule
{\fontsize{8pt}{10pt}\selectfont LLM-only}       & {\fontsize{8pt}{10pt}\selectfont $16.01 \pm 5.01$} & {\fontsize{8pt}{10pt}\selectfont $27.21 \pm 14.83$} & {\fontsize{8pt}{10pt}\selectfont \textasciitilde  0.65 TFLOPs}\\
{\fontsize{8pt}{10pt}\selectfont Naive RAG}  & {\fontsize{8pt}{10pt}\selectfont $3632 \pm 28.95$} & {\fontsize{8pt}{10pt}\selectfont $26.59 \pm 19.81$} & {\fontsize{8pt}{10pt}\selectfont \textasciitilde 22.98 TFLOPs}\\
{\fontsize{8pt}{10pt}\selectfont GraphRAG}   & {\fontsize{8pt}{10pt}\selectfont $9017\pm 2529$} & {\fontsize{8pt}{10pt}\selectfont $142.7 \pm 91.58$} & {\fontsize{8pt}{10pt}\selectfont \textasciitilde 58.57 TFLOPs}\\
\bottomrule
\end{tabular}
\vspace{-0.2cm}
\caption{Comparison of token utilization and inference computational cost among LLM-only, Naive RAG and GraphRAG (with default parameters) using a 3B LLM. The total computational cost, measured in TFLOPs, is estimated based on~\protect\cite{pope2023efficiently}.}
\label{tab:llm_comparison}
\vspace{-0.3cm}
\end{table}

At the same time, SLMs delpoy on edge have gained significant attentions, particularly when integrated with external knowledge sources, as seen in Naive RAG and RAG variants like GraphRAG~\cite{edge2024local}. As shown in Figure~\ref{fig:Combined_Model_Performance_and_Bottleneck}, leveraging SLMs for on-device inference can reduce delay and operational costs, making them suitable for dynamic industrial settings with varying QoS requirements. However, existing RAG implementations face critical challenges. Naive retrieval mechanisms can introduce irrelevant or misleading information, degrading model performance. Furthermore, cloud-based retrieval solutions, such as GraphRAG, add latency and increase token consumption, as shown in Table~\ref{tab:llm_comparison}. The overhead from retrieving large volumes of text as context can significantly raise the input-output token ratio, increasing TFLOPs consumption.
 

\begin{table}[t]
\small
\centering
\begin{tabular}{p{1.6cm}p{6cm}}
\toprule
\textbf{{\fontsize{8.5pt}{10pt}\selectfont  Query Type}} & \textbf{{\fontsize{8.5pt}{10pt}\selectfont Example Queries}} \\
\midrule
{\fontsize{8.5pt}{10pt}\selectfont Time} & 
{\fontsize{8pt}{10pt}\selectfont  Who won the 2022 World Cup?} \\
& {\fontsize{8pt}{10pt}\selectfont  Who are the candidates for the 2024 U.S. election?} \\
& {\fontsize{8pt}{10pt}\selectfont What are the breakthroughs in Tesla's latest autopilot?} \\
\midrule
{\fontsize{8.5pt}{10pt}\selectfont Location} & 
{\fontsize{8pt}{10pt}\selectfont What is the Alaska Permanent Fund Dividend?} \\
& {\fontsize{8pt}{10pt}\selectfont What are the Mardi Gras traditions in New Orleans? }\\
& {\fontsize{8pt}{10pt}\selectfont When is Vermont's maple syrup season, how to join?} \\
\bottomrule
\end{tabular}
\vspace{-0.2cm}
\caption{Examples of real-world queries with temporal and spatial variations, highlighting challenges in adapting retrieval models to dynamic user interests.}
\label{tab:time_space_variation}
\vspace{-0.3cm}
\end{table}

Moreover, real-world queries exhibit both temporal and spatial variations, posing further challenges for retrieval models. Table~\ref{tab:time_space_variation} illustrates how user interests fluctuate over time, with queries often reflecting evolving events, such as sports results, political elections, or technological advancements. Since LLMs are trained on datasets available up to a fixed cutoff date, even large models are inherently limited by outdated information. For instance, the powerful o1 model, without external updates, cannot recognize entities like o3 or R1. A retrieval system that fails to update its knowledge base frequently may struggle to provide accurate responses to time-sensitive queries. Similarly, spatial variations affect query relevance, as users from different locations seek information specific to their regional context, such as local policies, cultural traditions, or seasonal events. Without mechanisms for adapting to these dynamic changes, retrieval models risk conflating information from different regions or failing to retrieve relevant data, leading to incomplete or misleading responses. Existing RAG methods often lack the ability to dynamically adjust to these shifts in information demand, resulting in suboptimal retrieval quality and inference accuracy. In our experiments, we measured accuracy by comparing generated responses to ground truth using GPT-4o~\cite{langchain2023autoevaluator}.

To address these limitations, we propose EACO-RAG, a dynamic system designed to adapt to changing real-world conditions. Our approach integrates three core components: (1) dynamic knowledge updates at the edge, ensuring dataset relevance and accurate as contents of user interests evolve over time; (2) edge-assisted and collaboration, enabling retrieval across other relevant edge nodes rather than being limited to the local dataset, thereby expanding the scope of knowledge sets on the edge; and (3) an adaptive collaborative gating mechanism that selects the optimal retrieval strategy to meet various QoS demands. 

\section{EACO-RAG Design}

\subsection{Overview}
EACO-RAG integrates edge nodes with cloud resources, using lightweight models and adaptive retrieval strategies to reduce costs while maintaining high QoS. It features an edge-assisted RAG architecture that dynamically updates local knowledge bases, ensuring timely and relevant information. The system efficiently utilizes edge databases and SLMs for query answering. By enabling collaboration among edge nodes and incorporating a collaborative gating mechanism, EACO-RAG improves scalability and adaptability, allowing seamless operation across diverse real-world scenarios.

\subsection{Why GraphRAG}
A key enhancement of EACO-RAG is its integration with GraphRAG, a structured retrieval framework that improves accuracy by maintaining strong contextual relationships among knowledge elements. Unlike conventional RAG methods, GraphRAG organizes information into nodes, edges, and communities. Nodes represent discrete knowledge units, edges capture relationships, and communities group semantically related concepts. This structure enables more precise and context-aware retrieval, making GraphRAG particularly effective for multi-hop reasoning and knowledge-intensive tasks. However, its extended retrieval time and reliance on long-context processing make it more suitable for cloud deployment. Running GraphRAG directly on edge devices would be computationally restrictive and inefficient in utilizing edge resources.

EACO-RAG addresses these challenges by selectively extracting and distributing relevant knowledge from GraphRAG to the edge. This balances retrieval efficiency and contextual depth. The system dynamically updates the most relevant data chunks from GraphRAG communities, enabling localized knowledge adaptation without maintaining a full graph structure. Strong intra-community alignment in GraphRAG ensures that even lightweight mechanisms, like Naive RAG, operate with well-structured and semantically coherent data. This reduces ambiguity in concept interpretation and mitigates confusion from polysemous terms. For example, the term “model” may refer to a fashion model, a scientific model, or a prototype. By anchoring retrieval within relevant communities, EACO-RAG ensures that each edge node receives only the most contextually appropriate data. This minimizes definitional ambiguity while optimizing retrieval efficiency on the edges.

\subsection{System Design}
\small
\begin{figure}[t]
\centering
\vspace{-0.3cm}
\includegraphics[width=0.47\textwidth]{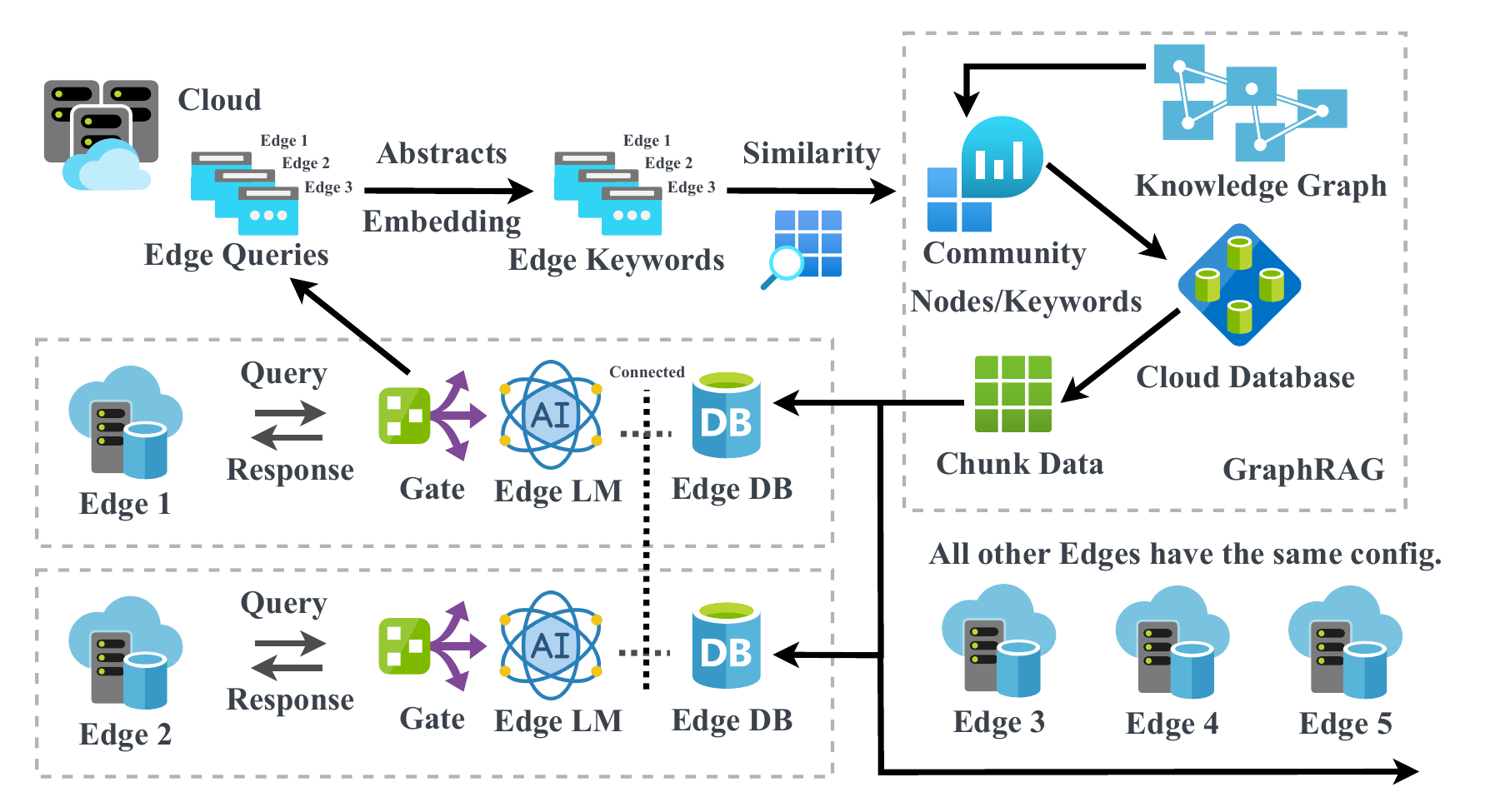}
\captionsetup{width=1\linewidth}
\vspace{-0.4cm}
\caption{Workflow of the EACO-RAG system design.}
\vspace{-0.3cm}
\label{fig:system_architecture}
\end{figure}

As illustrated in Figure~\ref{fig:system_architecture}, the cloud periodically collects and processes queries from users across various edge nodes, maintaining a knowledge graph that organizes nodes and communities based on evolving information trends. These nodes serve as keywords to extract relevant data chunks from recent queries at each edge location, allowing the system to capture both spatial and temporal shifts in user interests. The extracted data is then dynamically distributed to the appropriate edge nodes, reducing the dependency on large local databases while mitigating retrieval costs and resource overhead.

During query processing, EACO-RAG dynamically selects the optimal retrieval source by evaluating the overlap between an incoming query and the keyword-indexed knowledge at each edge node. This allows retrieval to extend beyond a single location, incorporating relevant insights from nearby nodes rather than being restricted to the most locally popular data. As information needs evolve over time, edge databases are continuously updated, ensuring that responses remain contextually relevant across both spatial and temporal dimensions.

Recognizing the various complexity of queries, EACO-RAG distinguishes between those that can be efficiently processed at the edge and those requiring escalation. While lightweight models and edge databases efficiently handle straightforward queries, more complex or multi-hop reasoning tasks may exceed their capacity due to limited local resources. To address this, the system escalates complex queries to the cloud-based knowledge graph or high-parameter models, improving accuracy while maintaining responsiveness and adaptability across diverse real-world scenarios.

At the core of this framework is the collaborative gating mechanism, which optimizes query processing by balancing cost, accuracy, and delay. For simple, well-covered queries, the system relies on SLMs at the edge for fast responses. When queries demand deeper reasoning or involve less common topics, retrieval is progressively escalated to more comprehensive knowledge sources, ensuring a balance between computational efficiency and response quality. For complicated queries that also demands high precision and reduced delay, additional cloud resources are allocated to execute large-parameter language models.

To optimize the balance between QoS and cost-effectiveness, EACO-RAG formulates decision-making as a contextual multi-armed bandit problem. Safe Online Bayesian Optimization continuously refines decision policies based on real-time query distribution and retrieval performance. By dynamically coordinating retrieval and inference across edge and cloud resources, EACO-RAG effectively reduces operational costs while ensuring accurate and timely responses, making it well-suited for dynamic and evolving application environments.

\section{Modelling and Optimization}

\subsection{ Modelling with Contextual Multi-Armed Bandit}

Our goal is to minimize operational costs, including resource and time costs. Resource costs come from model inference and retrieval, based on input and output tokens. Time costs account for network and loading delays. The system must satisfy two QoS constraints: (i) accuracy must exceed \( {QoS}_{\min}^\rho \), and (ii) response time must stay within \( {QoS}_{\max}^h \).

\textbf{Context}: At each time step $t$, the context is represented as $c_t := [d_t, s_t, q_t] \in \mathcal{C}$. Here, $d_t$ denotes network delays, which include both cloud and edge delays, helping assess network availability. The term $s_t$ consists of two values: the highest keyword overlap ratio between the query and the edge datasets, along with the corresponding edge dataset. The query complexity $q_t$ is represented as a set that captures whether the query requires single-hop or multi-hop reasoning, its length, and the number of entities it contains~\cite{yang2018hotpotqa}.

\textbf{Control Policies}: The system’s control policy at time $t$ is denoted as $x_t := [r_t, g_t] \in \mathcal{X}$. Here, $r_t$ selects the retrieval source, which can be none, edge-assisted naive retrieval, or cloud knowledge graph-based retrieval. The term $g_t$ determines the response generation location, either local SLM or cloud LLM. These policies dynamically adapt to real-time context to optimize cost while maintaining accuracy and delay constraints.

The total cost function is defined as:
\begin{equation}
u_t(c_t, x_t) = \delta_1 \cdot u_{\text{r}}^t(c_t, x_t) + \delta_2 \cdot u_{\text{d}}^t(c_t, x_t),
\end{equation}
where $\delta_1$ and $\delta_2$ weight resource and time costs. $u_{\text{r}}^t(c_t, x_t)$ represents computational costs, while $u_{\text{d}}^t(c_t, x_t)$ accounts for time costs. For ease of parameter adjustments, we unify the unit of resource and time costs by scaling the time cost with the peak TFLOPs of different GPUs depending on $c_t$ and $x_t$ as shown in Table~\ref{table:gpu_cost}, which turn out to also better reflect real-world situations as the time cost is usually minimal for edge devices but significant for cloud computing.

\begin{table}[t]
\centering
{\fontsize{8.5pt}{10pt}\selectfont
\setlength{\tabcolsep}{10pt} 
\begin{tabular}{lcc}
\toprule
\textbf{GPU Model} & \textbf{FP64 (Double Precision)} \\
\midrule
NVIDIA GeForce RTX 4090 & 1.29 TFLOPS\\
NVIDIA Tesla P100 & 4.70 TFLOPS\\
NVIDIA Tesla V100 & 7.80 TFLOPS\\
NVIDIA A100 Tensor Core & 9.70 TFLOPS\\
NVIDIA H100 Tensor Core & 60.00 TFLOPS\\
\bottomrule
\end{tabular}
\vspace{-0.2cm}
\caption{Double-precision (FP64) peak performance (in TFLOPS) of various server-side NVIDIA GPUs.}
\label{table:gpu_cost}
}
\vspace{-0.5cm}
\end{table}

The optimization problem is formulated as: 
\begin{equation}
\begin{aligned}
&\min_{\{x_t\}_{t=1}^T} \sum_{t=1}^T u_t(c_t, x_t) \\
&\text{s.t.} \quad \rho_t(c_t, x_t) \geq {QoS}_{\min}^\rho, \quad \forall t \leq T, \\
&\quad \quad h_t(c_t, x_t) \leq {QoS}_{\max}^h, \quad \forall t \leq T,
\end{aligned}
\end{equation}
where \(T\) is the total number of decision steps. $\rho_t(c_t, x_t)$ denotes answer accuracy and $h_t(c_t, x_t)$ represents overall delay at step \(t\). ${QoS}_{\min}^\rho$ and ${QoS}_{\max}^h$ denote the minimum accuracy and maximum delay, respectively. The goal is to minimize total cost while meeting accuracy and delay constraints at each step.

This formulation achieves cost efficiency while meeting the essential QoS constraints of accuracy and delay. The QoS constraints can be adjusted to suit different scenarios and applications, enabling various gate instances to address diverse requirements. In addition, the system adapts to changing contexts through online learning, which supports real-time decision-making. In the following subsection, we present our Safe Online Bayesian Optimization solution, which further details how the collaborative gate mechanism makes adaptive decisions in real time.

\begin{algorithm}[ht]
{\fontsize{8pt}{10pt}\selectfont
\caption{\fontsize{9.3pt}{15pt}\selectfont Collaborative Gating SafeOBO Algorithm}
\label{alg:SafeOBO}
\SetAlgoLined
\textbf{Inputs:} Control space $\mathcal{X}$, Safe seed set $S_0$, kernel $k$, ${QoS}_{\min}^\rho$ (minimum accuracy), ${QoS}_{\max}^h$ (maximum delay), exploration parameter $\beta$, cost weights $\delta_1$, $\delta_2$

\textbf{Initialization:} $Z_0 = \emptyset, y_0^{(0)} = \emptyset, y_0^{(1)} = \emptyset, y_0^{(2)} = \emptyset$\;

\For{$t = 1, \dots, T_0$ (Exploration phase)}{
    \textbf{Observe} context $c_t$\;
    \textbf{Randomly select} $x_t$ from $\mathcal{X}$ (warm-up step)\;
    
    \textbf{Observe} $h_t(c_t, x_t)$ (response time), $\rho_t(c_t, x_t)$ (accuracy), $u_{\text{r}}^t(c_t, x_t)$ (resource cost), $u_{\text{d}}^t(c_t, x_t)$ (delay cost)\;
    
    \textbf{Compute total cost:}
    \quad $u_t(c_t, x_t) = \delta_1 \cdot u_{\text{r}}^t(c_t, x_t) + \delta_2 \cdot u_{\text{d}}^t(c_t, x_t)$\;
    
    \textbf{Update} GP posteriors: \\
    \quad $y_t^{(0)} \gets y_{t-1}^{(0)} \cup u_t(c_t, x_t)$ (update cost posterior)\;
    \quad $y_t^{(1)} \gets y_{t-1}^{(1)} \cup \rho_t(c_t, x_t)$ (update accuracy posterior)\;
    \quad $y_t^{(2)} \gets y_{t-1}^{(2)} \cup h_t(c_t, x_t)$ ( update response time posterior)\;
}

\For{$t = T_0 + 1, \dots, T$ (Exploitation phase)}{
    \textbf{Observe} context $c_t$\;
    \textbf{Compute} $\mu_{t-1}^{(i)}(c_t, x)$ and $\sigma_{t-1}^{(i)}(c_t, x)$ for all $i = 0, 1, 2$, using the posterior from the previous iteration\;

    \textbf{Estimate the safe set:} \\
    \quad $S_t = S_0 \cup \{x \in \mathcal{X} \mid \mu_t^{(1)}(c_t, x) - \beta \sigma_t^{(1)}(c_t, x) \geq {QoS}_{\min}^\rho$\\
    \quad $\land \mu_t^{(2)}(c_t, x) + \beta \sigma_t^{(2)}(c_t, x) \leq {QoS}_{\max}^h\}$\;

    \textbf{Select} $x_t = \arg\min_{x \in S_t} \mu_t^{(0)}(c_t, x) - \beta_t \sigma_t^{(0)}(c_t, x)$\;
    
    \textbf{Observe} $h_t(c_t, x_t)$ (response time), $\rho_t(c_t, x_t)$ (accuracy), $u_{\text{r}}^t(c_t, x_t)$ (resource cost), $u_{\text{d}}^t(c_t, x_t)$ (delay cost)\;
    
    \textbf{Compute total cost:}
    \quad $u_t(c_t, x_t) = \delta_1 \cdot u_{\text{r}}^t(c_t, x_t) + \delta_2 \cdot u_{\text{d}}^t(c_t, x_t)$\;
    
    \textbf{Update} GP posteriors: \\
    \quad $y_t^{(0)} \gets y_{t-1}^{(0)} \cup u_t(c_t, x_t)$ (update cost posterior)\;
    \quad $y_t^{(1)} \gets y_{t-1}^{(1)} \cup \rho_t(c_t, x_t)$ (update accuracy posterior)\;
    \quad $y_t^{(2)} \gets y_{t-1}^{(2)} \cup h_t(c_t, x_t)$ (update response time posterior)\;
}
}
\end{algorithm}

\subsection{Safe Online Bayesian Optimization}
\label{sec:solution}
We propose a Bayesian online optimization approach for decision-making in dynamic environments, outlined in Algorithm~\ref{alg:SafeOBO}. It uses Gaussian Processes (GPs) to model system correlations and quantify uncertainty, enabling adaptive optimization while reducing cost and maintaining performance and delay constraints.

\textbf{Function Approximation.} The algorithm use GPs to estimate cost and constraint functions, following established methods~\cite{williams2006gaussian,duvenaud2014automatic}. Each function is modeled as $\text{GP}(\mu(x), k(x, x'))$, where $\mu(x)$ is the mean function and $k(x, x')$ captures covariance. Based on observed data, the algorithm iteratively update posterior distributions for cost, accuracy, and delay to refine estimates.

\textbf{Safe Set Identification.} The safe set $S_t$ consists of control policies satisfying system constraints at time $t$:
\begin{equation}
\begin{aligned}
S_t = \{x \in \mathcal{X} \mid \mu_t^{(1)}(c_t, x) - \beta \sigma_t^{(1)}(c_t, x) \geq {QoS}_{\min}^\rho \\
    \quad \land  \mu_t^{(2)}(c_t, x) + \beta \sigma_t^{(2)}(c_t, x) \leq {QoS}_{\max}^h \}.
\end{aligned}
\end{equation}
Here, $\mu_t^{(i)}(c_t, x)$ and $\sigma_t^{(i)}(c_t, x)$ represent the GP-predicted mean and uncertainty for function $i$ (accuracy or delay). The parameter $\beta$ adjusts the confidence bound, balancing exploration and safety.

\textbf{Exploration and Exploitation.} The algorithm begins with an exploration (warm-up) phase, randomly selecting decisions to build an initial dataset of cost, accuracy, and response time. Once sufficient data is collected, the system shifts to an exploitation phase, optimizing decisions within the safe set to minimize total cost:
\begin{align}
x_t = \arg\min_{x \in S_t} \mu_t^{(0)}(c_t, x) - \beta_t \sigma_t^{(0)}(c_t, x).
\end{align}
Here, $\mu_t^{(0)}(c_t, x)$ and $\sigma_t^{(0)}(c_t, x)$ denote the predicted mean and uncertainty of the cost function. 

This strategy ensures cost efficiency while meeting accuracy and delay constraints. As new data continuously refine decision-making and update the GP models, the system adapts to changing contexts in real time, further enhancing its overall adaptability.

\section{Prototype Implementation}
\label{sec:implementation}


Our implementation employs a dual-layer architecture consisting of cloud and edge components. At the edge, a RTX 4090 GPU simulates multiple edge nodes running the SLM, local RAG, and EACO-RAG. The cloud is simulated using an A800 GPU emulating an 8$\times$H100 setup, supporting a 72B model with graphRAG-based retrieval and achieving approximately 1-second query response time. To maintain relevant local knowledge, the system dynamically updates a repository of 1,000 local data chunks, triggering updates when the cloud accumulates 20 new QA pairs. Keywords extracted from these queries guide updates via a first-in-first-out (FIFO) policy, ensuring efficient storage and retrieval performance.

Recall that beyond local inference, EACO-RAG also incorporates an edge-assisted retrieval mechanism, selecting datasets from other edge nodes when local coverage is insufficient. The system evaluates the overlap ratio, defined as the proportion of query keywords present in the target dataset, to determine the most relevant retrieval edge data chunks. To identify valid query keywords, a lightweight semantic embedding model ('all-MiniLM-L6-v2') is used to map the query to related keywords, considering those with a similarity score above 50\% as valid matches. A similar strategy is applied to GraphRAG-based adaptive knowledge updates, where recent edge queries are processed through the embedding model to identify relevant keywords in GraphRAG. The system then selects the top-$k$ communities containing the highest number of similar keywords or nodes and distributes up to 500 data chunks from these communities to the edge.

Additionally, the collaborative gating mechanism is also implemented to dynamically select the optimal inference path—ranging from local small-model inference to full cloud-based retrieval with the 72B model—based on query complexity and system constraints. For domain-specific retrieval, knowledge graphs are constructed from the corresponding knowledge sources associated with the datasets used for performance evaluation (i.e., 139 Wikipedia pages for Wiki QA dataset and the seven canonical Harry Potter books for Harry Potter QA dataset as specified in next section). This adaptive approach optimizes response quality while balancing computational efficiency and resource utilization.

\vspace{-0.15cm}
\section{Performance Evaluation}
\label{sec:evaluation}

\begin{table*}[ht]
{\fontsize{8.5pt}{10pt}\selectfont
    \centering
    \begin{tabular}{lccc ccc}
        \toprule
        & \multicolumn{3}{c}{\textbf{Wiki QA}} & \multicolumn{3}{c}{\textbf{Harry Potter QA}} \\
        \cmidrule(lr){2-4} \cmidrule(lr){5-7}
        & \textbf{Accuracy} (\%) & \textbf{Delay} (s) & \textbf{Cost} (TFLOPs) & \textbf{Accuracy} (\%) & \textbf{Delay} (s) & \textbf{Cost} (TFLOPs) \\
        \midrule
        3b LLM-only & 28.72 & 0.30 ± 0.07 & 0.60 ± 0.16 & 31.69 & 0.31 ± 0.08 & 0.65 ± 0.20 \\
        3b LLM+Naive RAG & 61.57 & 0.88 ± 0.11 & 23.10 ± 0.34 & 52.54 & 1.00 ± 0.18 & 23.62 ± 0.38 \\
        3b LLM+GraphRAG & 76.01 & 3.01 ± 1.21 & 60.02 ± 17.45 & 63.47 & 2.82 ± 1.32 & 58.99 ± 16.69 \\
        72b LLM+GraphRAG & 94.39 & 0.97 ± 0.64 & 711.43 ± 309.52 & 77.12 & 1.03 ± 0.84 &  739.79 ± 402.18 \\
        \midrule
        \textbf{EACO-RAG (Cost-Efficient)} & \textbf{94.92} & 1.27 & \textbf{109.40} & \textbf{78.00} & 1.74 & \textbf{139.43} \\
        \textbf{EACO-RAG (Delay-Oriented)} & 94.17 & \textbf{0.75} & 247.03 & 76.28 & \textbf{0.79} & 496.19 \\
        \bottomrule
    \end{tabular}
    \vspace{-0.2cm}
    \caption{Performance comparison of EACO-RAG and baseline approaches under our dual-layer (edge–cloud) architecture on Wiki QA and Harry Potter QA tasks, highlighting trade-offs among accuracy, delay, and cost.}
    \label{tab:qa_comparison}
    \vspace{-0.3cm}
}
\end{table*}

\subsection{Datasets Selection}

To evaluate the effectiveness of EACO-RAG, we configure two distinct datasets. The first dataset, referred to as Wiki QA, is based on Wikipedia and is designed to represent general-domain questions. The second dataset focuses on the Harry Potter series and is intended to emulate more specialized, industrial scenarios.

\textbf{Wiki QA Dataset:} We selected 139 popular Wikipedia pages from the Natural Questions (NQ) dataset~\cite{kwiatkowski2019natural}, each containing more than 10 question-answer pairs. To enhance the dataset's complexity and diversity, we integrated additional QA pairs from TriviaQA~\cite{joshi2017triviaqa} and HotpotQA~\cite{yang2018hotpotqa} corresponding to these pages, resulting in a comprehensive dataset of 571 QA pairs.

\textbf{Harry Potter QA Dataset:} Sourced from~\cite{saracandu_harrypotter_trivia}, this dataset contains 1,180 high-quality question-answer pairs covering various aspects of the Harry Potter series. We filtered the dataset to retain more reliable pairs. Compared to the Wiki QA dataset, the Harry Potter QA dataset features more challenging questions that require specific background knowledge, effectively simulating complex industrial scenarios.

By employing these two datasets, we aim to rigorously test EACO-RAG's performance across both general and specialized domains, demonstrating its versatility and effectiveness in handling a range of query complexities.

\begin{table}[t]
{\fontsize{8.5pt}{10pt}\selectfont
    \centering
    \begin{tabular}{lccc}
        \toprule
        \textbf{Warm-up Steps}& \textbf{Accuracy} (\%) & \textbf{Delay} (s) & \textbf{Cost} (TFLOPs) \\
        \midrule
        \multicolumn{4}{c}{Wiki QA} \\
        \midrule
        \textbf{EACO-RAG-300} & \textbf{94.92} & 1.27 & \textbf{109.40}  \\
        \textbf{EACO-RAG-200} & 89.66 & 1.26 & 140.06  \\
        \textbf{EACO-RAG-100} & 87.22 & 1.49 & 346.29 \\
        \midrule
        \multicolumn{4}{c}{Harry Potter QA} \\
        \midrule
        \textbf{EACO-RAG-500} & \textbf{78.00} & 1.74 & \textbf{139.43} \\
        \textbf{EACO-RAG-300} & 77.35 & 1.12 & 402.19 \\
        \textbf{EACO-RAG-100} & 74.44 & 1.31 & 511.60 \\
        \bottomrule
    \end{tabular}
    \vspace{-0.2cm}
    \caption{Effect of different warm-up steps on EACO-RAG’s gating decisions for Wiki QA and Harry Potter QA, showing how initial exploration influence accuracy, delay, and computational cost.}
    \vspace{-0.3cm}
    \label{tab:warm_up_comparsion}
}
\end{table}

\subsection{Overall Performance}
To evaluate the effectiveness of EACO-RAG, we compare EACO-RAG with several retrieval and generation baselines, including standalone SLMs, naive RAG-based edge retrieval, and cloud-based GraphRAG retrieval using both 3B and 72B parameter LLMs, as shown in Table~\ref{tab:qa_comparison}. EACO-RAG dynamically selects among these strategies based on query context and system constraints. We evaluate EACO-RAG under two different settings. In a cost-efficient setting, where delays up to 5s are acceptable, the gate prioritizes lower costs by favoring local inference and edge retrieval. In a delay-oriented setting, where responses must be under 1s, the gate relies more on cloud-based inference when network conditions allow, leading to higher computational costs.

Results demonstrate that EACO-RAG effectively leverages edge-assisted retrieval to significantly reducing costs while maintain accuracy and delay. Under relaxed latency constraints with strict accuracy requirements, it achieves performance comparable to a 72B LLM+GraphRAG system while reducing costs by 84.6\% on the Wiki QA dataset. Similarly, on the more challenging Harry Potter dataset, it attains similar accuracy with an 81.2\% cost reduction. In the delay-oriented setting, cost reductions are 65.3\% for WikiQA and 32.9\% for Harry Potter QA. This strong performance is attributed to EACO-RAG’s ability to dynamically select integrated strategies that optimize the trade-off between cost and QoS.

Table~\ref{tab:qa_example_summary} further takes a closer look at two examples on how the collaborative gating mechanism selects the optimal answer path for each query. In Question 1, the query is recognized as a simple single-hop problem, and the context indicates that the edge dataset fully covers the relevant entities with low delay; consequently, the gate selects the corresponding edge dataset and local 3B SLM. In contrast, in Question 2 the query is identified as a complex multi-hop problem based on contextual cues, and the context reveals that edge-assisted retrieval is insufficient—thus, the gate opts for cloud-based collaborative retrieval and generation.

These results confirm the effectiveness of our method, demonstrating that EACO-RAG can significantly lower costs across different constraints while maintaining robust performance.


\subsection{Impact of Warm-up Steps}


The number of warm-up steps ($T_0$) in EACO-RAG significantly influences the gate’s strategy selection. An increased amount of warm-up data enhances the gate's contextual understanding, enabling the system to better differentiate between queries that can be answered locally and those requiring GraphRAG or cloud-based LLM inference. We evaluated various warm-up sizes and measured their impact on accuracy, delay, and cost. As shown in Table~\ref{tab:warm_up_comparsion}, for Wiki QA—with relatively simple queries—the gate begins favoring edge-based responses at 100 warm-up steps and effectively distinguishes query by 300 steps. In contrast, the Harry Potter dataset, which contains more specialized and context-dependent queries, requires more warm-up data before the system can shift more queries to edge-based inference to reduce costs.

\subsection{Comparison of Different SLMs}

\begin{table}[t]
{\fontsize{8.5pt}{10pt}\selectfont
\centering
\begin{tabular}{lccc}
    \toprule
    \textbf{Model} & \textbf{Accuracy} (\%) & \textbf{Delay} (s) & \textbf{Cost} (TFLOPs) \\
    \midrule
    Qwen2.5 7B   & 94.57 & 1.48 & 93.83  \\
    Qwen2.5 3B   & 94.92 & 1.27 & 109.40\\
    llama3.2 3B  & 93.35 & 1.07 & 272.72  \\
    Qwen2.5 1.5B & 91.42 & 0.95 & 167.67  \\
    \bottomrule
\end{tabular}
\vspace{-0.2cm}
\caption{EACO-RAG performance of various SLMs on Wiki QA.}
\label{tab:sml_comparision}
\vspace{-0.5cm}
}
\end{table}

In our previous experiments, we deployed a uniform 3B SLM at the edge for adaptability and scalability across edge devices with varying computational capabilities. To further evaluate EACO-RAG with SLMs of varying sizes and origins, we tested several open-source SLMs on the WikiQA dataset, including Qwen2.5 7B, Qwen2.5 1.5B, and llama3.2 3B. As shown in Table~\ref{tab:sml_comparision}, replacing the edge SLM with Qwen2.5 7B reduced overall cost despite its higher computational expense, as the gate identified more queries that could be resolved at the edge. In contrast, using Qwen2.5 1.5B significantly lowered delay but increased cost, indicating that more queries were escalated to cloud-based retrieval. 

Notably, the llama3.2 3B model underperformed compared to Qwen2.5 3B. This difference is due to distinct training approaches. Qwen2.5 3B benefits from large-scale pre-training and targeted fine-tuning~\cite{qwen2.5_llm_blog}, which enhances its knowledge and contextual reasoning~\cite{kaplan2020scaling}. Llama3.2 3B relies on pruning and distillation~\cite{meta_llama_3.2_blog}, resulting in a lighter model with faster inference but reduced reasoning ability~\cite{jiao2019tinybert}. As contextual understanding is critical because EACO-RAG relies on contextual understanding for retrieval-augmented generation in both edge-assisted and cloud-based methods, in this case, LLaMA3.2 underperforms compared to Qwen2.5 on EACO-RAG.


\begin{table*}[ht]
\centering
{\fontsize{8.5pt}{10pt}\selectfont
\begin{tabular}{@{}l p{1.8\columnwidth}@{}}
\toprule
\textbf{Question 1} & What is the name of the spell used to unlock doors? \\[4pt]
\midrule
\textbf{Process} & Context:\{Single-hop; 15 tokens; 3 entities (spell, unlock, door); Edge4:[100\% match, 20\,ms delay]; Cloud:[300\,ms delay]\} \(\Rightarrow\) Gate \(\Rightarrow\) Decision:\{Edge4 dataset + 3B SLM\} \\[4pt]
\midrule
\textbf{Output} & Alohomora (Correct) \\[4pt]
\midrule
\addlinespace[4pt]
\midrule
\textbf{Question 2} & What impact does Harry’s friendship with Hermione have on his understanding of empathy and compassion? \\[4pt]
\midrule
\textbf{Process} & Context:\{Multi-hop; 21 tokens; 4 entities (Harry, Hermione, empathy, compassion); Edge6:[25\% match, 32\,ms delay]; Cloud:[350\,ms delay]\} \(\Rightarrow\) Gate \(\Rightarrow\) Decision:\{Cloud GraphRAG + 72B LLM\} \\[4pt]
\midrule
\textbf{Output} & Harry’s friendship with Hermione deepens his empathy and compassion through her intellect, support, and loyalty, helping him consider others' perspectives and act with kindness. (Correct) \\
\bottomrule
\end{tabular}
\vspace{-0.2cm}
\caption{Illustrative examples of QA processing in EACO-RAG using the collaborative gate mechanism.}
\label{tab:qa_example_summary}
}
\vspace{-0.3cm}
\end{table*}


\subsection{Ablation Study and Hyperparameter Analysis}

\begin{figure}[t]
\vspace{-0.5cm}
\small
  \centering
  \subfigure{
    \includegraphics[width=0.45\linewidth]{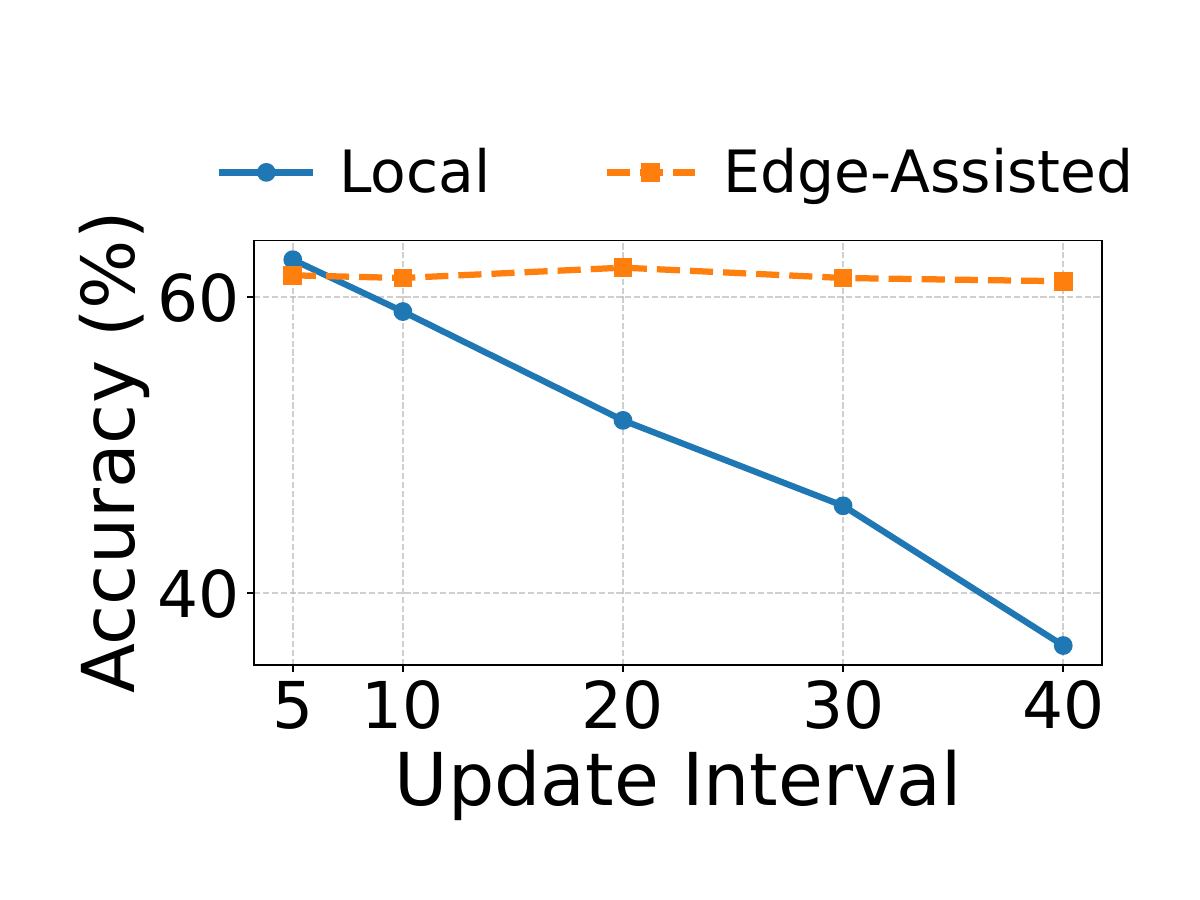}
    \label{Fig:Update_Step_Accuracy_Comparison}
  }
  \subfigure{
    \includegraphics[width=0.45\linewidth]{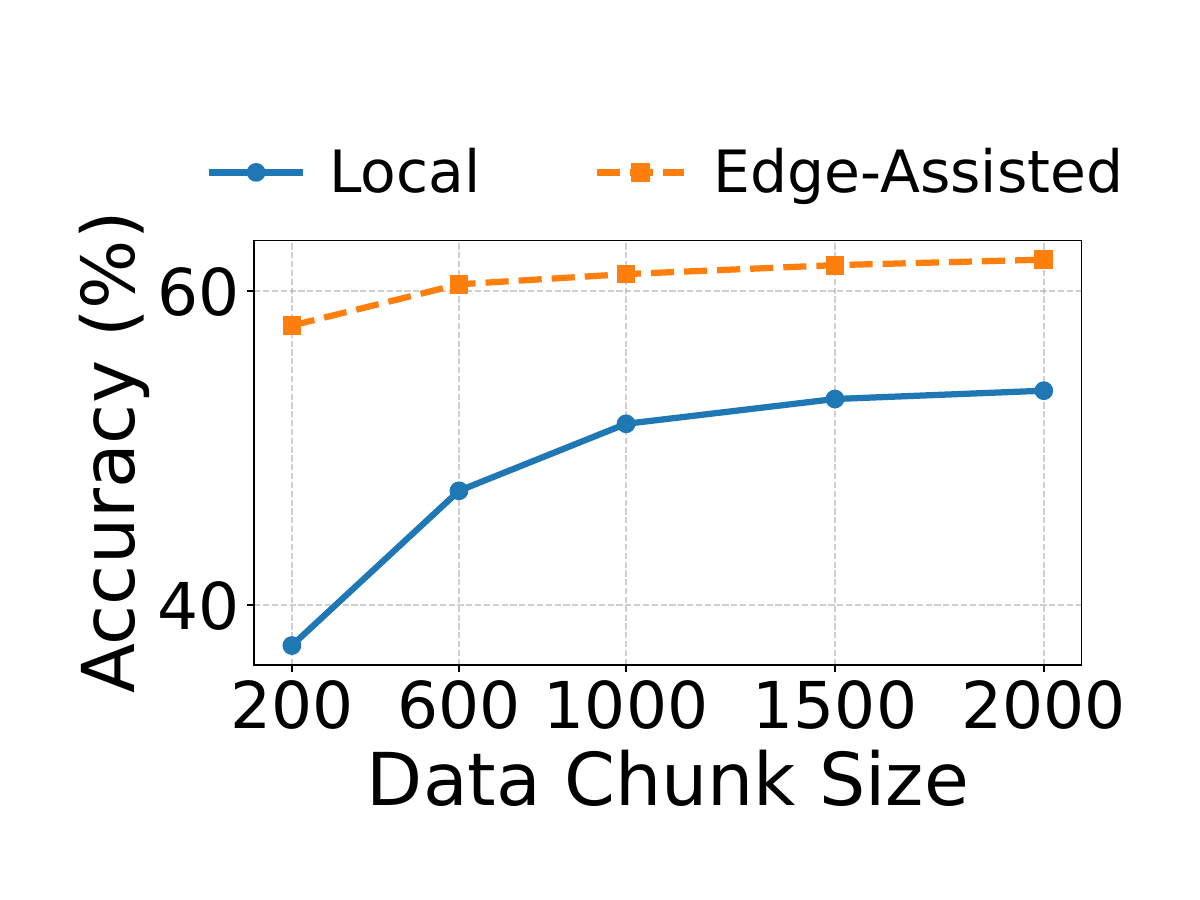}
    \label{Fig:Chunk_size_Accuracy_Comparison}
  }
  \captionsetup{width=1\linewidth} 
  \vspace{-0.5cm}
  \caption{Accuracy comparison under different hyperparameter settings in the ablation study. \textbf{Left:} shows how varying the local adaptive update trigger interval influences accuracy. \textbf{Right:} illustrates the effect of different chunk sizes in the edge dataset.}
  \vspace{-0.39cm} 
\end{figure}

In this study, we remove the collaborative gating mechanism and exclude cloud retrieval and generation. This isolates the effects of adaptive knowledge update and edge-assisted retrieval on accuracy. Without cloud retrieval, delay and cost differences between local and edge responses become negligible. We focus solely on evaluating the accuracy impact of these two components under various hyperparameter settings.

We first examine the effect of the local adaptive update trigger interval. We compare naive RAG relying solely on the local update database with and without edge-assisted retrieval. As shown in Figure~\ref{Fig:Update_Step_Accuracy_Comparison}, when relying solely on local dataset, the update interval has a strong impact on accuracy. Incorporating edge-assisted retrieval reduces this sensitivity. Frequent local updates may even outperform edge-assisted retrieval due to stronger contextual correlations from the knowledge graph, but they consume more resources. We also study the impact of local chunk size. Figure~\ref{Fig:Chunk_size_Accuracy_Comparison} indicates that larger chunk sizes yield higher accuracy for both methods. With edge-assisted retrieval, convergence occurs at 600 data chunks; without edge-assisted, over 1000 data chunks are needed. However, on resource-constrained edge devices, a larger chunk size adds to the retrieval burden. In summary, more frequent updates and larger chunk sizes improve accuracy but increase resource overhead, while edge-assisted retrieval reduces sensitivity to both parameters. These findings also demonstrate that adaptive knowledge update and edge-assisted retrieval effectively address issues of outdated datasets and limited diversity from region-specific edge datasets.

\vspace{-0.2cm}
\section{Related Work}
\label{sec:related_work}

\textbf{Retrieval-Augmented Generation Enhancements.} RAG improves language models by integrating relevant text from knowledge bases~\cite{lewis2020retrieval}. Extensions such as Adaptive RAG~\cite{jeong2024adaptive}, Corrective RAG~\cite{yan2024corrective}, and Self-RAG~\cite{asai2023self} addressed limitations in retrieval strategies and query complexity by adapting operations based on query difficulty. Mallen et al.~\cite{mallen-etal-2023-trust} classified query complexity via entity frequency to guide binary decisions on retrieval sufficiency, while Qi et al.~\cite{qi-etal-2021-answering} employed fixed operations (retrieval, reading, reranking) that require specialized LM training. Despite these advances, traditional methods often rely on centralized frameworks. In contrast, EACO-RAG leverages distributed edge computing to dynamically update local databases, reducing delays and communication overhead.

\textbf{Edge-Enabled Cost Optimization and Resource Allocation.} Reducing LLM deployment costs has attracted significant research attention. Techniques such as model quantization~\cite{xiao2023smoothquant,park2023lut}, pruning~\cite{ma2024llm}, and distillation enabled smaller models to mimic larger ones, while caching strategies~\cite{stogiannidis2023cache,zhu2023optimalcaching,gill2024privacy,li2024scalm} and key-value state reuse~\cite{liu2024cachegen,yao2024cacheblend} further lowered computational expenses. Additionally, approaches like FrugalGPT~\cite{chen2023frugalgpt} and model multiplexing~\cite{bang2023gptcache,kim2023biglittle} dynamically adjusted model size based on query complexity. On the edge computing front, research has focused on optimizing resource allocation by balancing delay, energy, and processing power through task offloading, resource scheduling~\cite{naouri2021novel}, and edge-cloud collaboration~\cite{gu2023ai,xiong2020resource}. EACO-RAG offers a holistic solution that integrates both edge and cloud resources and introduces inter-node collaboration for dynamic knowledge sharing and updating, thereby optimizing retrieval and generation while minimizing operational costs.

\vspace{-0.2cm}
\section{Further Discussion}
The limitations of EACO-RAG stem from its constrained experimental scope. In our study, the collaborative gating mechanism only selects among four retrieval and inference strategies. In real-world applications, a broader range of adaptive strategies may emerge, further enhancing efficiency and accuracy. It is also interesting to have more large-scale studies and deployments in the future to fully assess the system's capabilities. In addition, the current gating design in our prototype implementation relies on general heuristics. Yet, query features can vary greatly for different datasets. Therefore, in practical industrial settings, the gate could further incorporate dataset-specific characteristics to provide richer contextual cues. This integration can better improve decision-making, adaptability, and optimization across diverse environments.

\vspace{-0.2cm}
\section{Conclusion}
\label{sec:conclusion}

This paper presented EACO-RAG, an edge-assisted and collaborative RAG system that uses adaptive knowledge update to compensate query evolving across both region and time, and minimizes cost while maintaining high accuracy and low delay. By dynamically selecting among local, edge-assisted, and cloud-based strategies, our method achieved accuracy comparable to cloud-based RAG at significantly lower inference costs. Our experiments showed that our approach delivered substantial gains across various parameter settings, datasets, and parameters, demonstrating its adaptability and scalability. Future work will focus on improving edge dataset knowledge management for graph-based retrieval, further optimizing and validating the system in larger-scale scenarios.

\clearpage

\appendix





\bibliographystyle{named}
\bibliography{ijcai25}

\end{document}